\documentclass[showpacs,twocolumn,prb]{revtex4-1}
\usepackage{amsfonts}
\usepackage{graphicx}
\usepackage{amsmath}
\usepackage{amssymb}
\usepackage{color}

\begin{document}

\title{Impact of spin-orbit coupling on the Holstein polaron}
\author{Zhou Li$^1$, L. Covaci$^2$, M. Berciu$^3$, D. Baillie$^{1*}$, F. Marsiglio$^1$}
\affiliation{
$^1$ Department of Physics, University of Alberta,
Edmonton, Alberta, Canada, T6G~2J1
\\
$^2$Departement Fysica, Universiteit Antwerpen, Groenenborgerlaan 171, B-2020 Antwerpen, Belgium
\\
$^3$Department of Physics and Astronomy, University of British Columbia, Vancouver, BC, Canada, V6T 1Z1
}

\begin{abstract}
We utilize an exact variational numerical procedure to calculate the
ground state properties of a polaron in the presence of a Rashba-like
spin orbit interaction. Our results corroborate with previous work
performed with the Momentum Average approximation and with weak coupling
perturbation theory. We find that spin orbit coupling increases the
effective mass in the regime with weak electron phonon coupling, and
decreases the effective mass in the intermediate and strong electron
phonon coupling regime. Analytical strong coupling perturbation theory
results confirm our numerical results in the small polaron regime. A
large amount of spin orbit coupling can lead to a significant lowering
of the polaron effective mass.
\end{abstract}


\date{\today}
\maketitle

\section{Introduction}

In much of condensed matter (magnetism excepted), the spin and orbital
components of an electron are treated as independent degrees of freedom.
Nonetheless, the non-relativistic approximation to the Dirac equation leads
directly to the so-called Thomas term in the effective Hamiltonian, which
can be written as a spin-orbit coupling term.\cite{sakurai67} This coupling
can play a significant role in the electronic structure of semiconductors
and metals, as documented, for example, in Ref. (\onlinecite{winkler03}).
More recently, interest has grown because of the burgeoning possibilities in
the so-called field of \emph{spintronics}, where the spin degree of freedom
is specifically exploited for potential applications.\cite{wolf01} Control
of spin will require coupling to the orbital motion, and hence spin-orbit
coupling may play a critical role in understanding and exploiting various
properties of such systems.

Spin-orbit coupling, as described by Rashba,\cite{rashba60} is expected to
be prominent in two dimensional systems that lack inversion symmetry,
including surface states. Many such systems have now been identified, among
which are, for example, surface alloys, Li/W(110),\cite{rotenberg99}
Pb/Ag(111),\cite{pacile06,ast07a} and Bi/Ag(111).\cite{ast07b} In all of these
systems the possibility of other interactions remains; in particular recent
work\cite{cappelluti07} has focussed on the electron-phonon interaction, in
the presence of Rashba spin-orbit interactions. In the first reference of
Ref. (\onlinecite{cappelluti07}), for example, the effective mass due to the
electron-phonon interaction was shown, in weak coupling, to be enhanced by
the spin-orbit interaction.

More recently, attention has focussed on the properties of a single electron
interacting with oscillator degrees of freedom \cite{holstein59} in the
presence of Rashba spin-orbit coupling.\cite{covaci09} These authors
utilized the so-called momentum average (MA) approximation\cite{berciu06} to
examine the properties of a single polaron also in the presence of
spin-orbit coupling, but for a tight-binding model. They found that the
effective mass generally decreases as a function of spin-orbit coupling, $V_{%
\mathrm{S}}$; however, in the weak electron-phonon coupling limit, there is
initially an increase in effective mass, in agreement with Cappelluti et al.
\cite{cappelluti07} The primary purpose of this work is to present exact
solutions to this problem, using Trugman's method,\cite{trugman90,bonca99}
along with some modified algorithms,\cite{li10} so that we can span the
entire parameter regime. It turns out that the MA method is fairly accurate
over the entire parameter range, except for low phonon frequency.

We also develop a strong coupling expansion, based on the Lang-Firsov
transformation,\cite{lang63} following Ref. (\onlinecite{marsiglio95}). As
in the straightforward Holstein model, strong coupling describes fairly well
the small polaron regime. Finally, the adiabatic limit of the Holstein model
with Rashba spin-orbit coupling has been described recently in Ref. (\onlinecite{grimaldi10}),
following Refs. (\onlinecite{lagendijk85}) and (\onlinecite{kabanov93}) for the simple Holstein
model. In the strict adiabatic limit Grimaldi
finds an intermediate state (large polaron) with the lowest energy, for
coupling strengths just below that required for small polaron formation, in
the presence of spin-orbit coupling (see Figs. 1 and 2 in Ref. (%
\onlinecite{grimaldi10})). Our search for this state will also be described
in the present work.

The paper is organized as follows. We first introduce the model of study;
following Ref. (\onlinecite{covaci09}) it is the Holstein model with
additional Rashba spin-orbit coupling, written for a tight-binding
formulation. We note some of the features of the non-interacting (with
respect to phonons) model. Unlike the continuum limit,\cite{cappelluti07}
there is not a singularity at the bottom of the band; however, for weak
spin-orbit coupling, a singularity remains very close by in energy, and
causes a significant enhancement in the density of states at the bottom of
the band. In Section III we present our numerical results, along with those
from the strong coupling expansion and from the MA approximation. As
mentioned above, the exact numerical results confirm the conclusions from
Ref. (\onlinecite{covaci09}). Finally, we examine the low phonon frequency
and intermediate electron-phonon and spin-orbit coupling regimes, where both
perturbative and MA approaches are suspect. We are unable to rule out the
presence of an intermediate phase completely, but find that its occurrence
is unlikely, once quantum fluctuations are included. We close with a summary.

\section{Model}

The standard formulation for spin-orbit interaction uses two different types
of electronic band structure. The first is free electron-like, which results
in parabolic bands,\cite{cappelluti07} and the second is tight-binding,
which results in a periodic momentum dependence. While it is essentially
always the case that the latter tends to the former for low electron
fillings, this is not quite true when a Rashba-type spin-orbit interaction term is present.
As shown in Ref.[\onlinecite{cappelluti07}], for
example, the ground state for a single electron consists of a degenerate
ring around the $\Gamma -$point. This results in an electronic density of
states with a square-root singularity at the bottom of the band. For a
tight-binding model, however, Covaci et al. \cite{covaci09} pointed out
that this is not the case. We will adopt a tight-binding
formulation here, and examine this difference more closely in the next sub-section.

To study the single polaron with spin-orbit interaction we use a
tight-binding Hamiltonian with Rashba-type spin-orbit interaction\cite{rashba60} and a
Holstein-type\cite{holstein59} electron-phonon interaction. In real space the Hamiltonian is:
\begin{eqnarray}
H &=&-t\sum_{<i,j>,\alpha =\uparrow \downarrow }(c_{i,\alpha }^{\dagger
}c_{j,\alpha }+c_{j,\alpha }^{\dagger }c_{i,\alpha })  \notag \\
&&+V_{S}\sum_{i,\alpha ,\beta }(ic_{i,\alpha }^{\dagger }\sigma _{x}^{\alpha
\beta }c_{i+\hat{y},\beta }-ic_{i,\alpha }^{\dagger }\sigma _{y}^{\alpha
\beta }c_{i+\hat{x},\beta }+h.c.)  \notag \\
&&-g\omega _{E}\sum_{i,s=\uparrow \downarrow }c_{i,s}^{\dagger
}c_{i,s}(a_{i}+a_{i}^{\dagger })+\omega _{E}\sum_{i}a_{i}^{\dagger }a_{i},
\end{eqnarray}%
where $c_{i,s}^{\dagger }$ ($c_{i,s}$) is the creation (annihilation) for an
electron at site $i$ with spin index $s$, $a_{i}^{\dagger
}$ ($a_{i}$) is the creation (annihilation) operator for a phonon
at site $i$, and $\sigma _{x}^{\alpha \beta },\sigma _{y}^{\alpha \beta }$
designate the $(\alpha,\beta)$ component of the usual Pauli matrices.
The sum over $i$ is over all sites in the
lattice, whereas $<i,j>$ means only nearest neighbor hopping is included.
Here, as the notation already suggests, we confine ourselves to nearest
neighbor hopping only. The energy scales are the hopping integral $t$, the
strength of the Rashba spin-orbit interaction, $V_{S}$, the coupling of the
electron to the oscillator degrees of freedom $g\omega _{E}$, and the
Einstein phonon frequency, $\omega _{E}$.
In what follows we write all energy scales in
terms of the hopping integral, $t$, which hereafter is set to unity. The
ground-state properties of the Holstein model in one and two dimensions near
the adiabatic limit have recently been studied in Refs. (\onlinecite{li10}) and (\onlinecite{alvermann10}).
Normally spin is not considered, since this ground state is degenerate with
respect to spin. As the Rashba spin-orbit interaction is turned on, however,
the two-fold degeneracy will be lifted.

\subsection{non-interacting model: ground state and effective mass}

\begin{figure}[tp]
\begin{center}
\includegraphics[height=2.1in,width=2.1in,angle=-90]{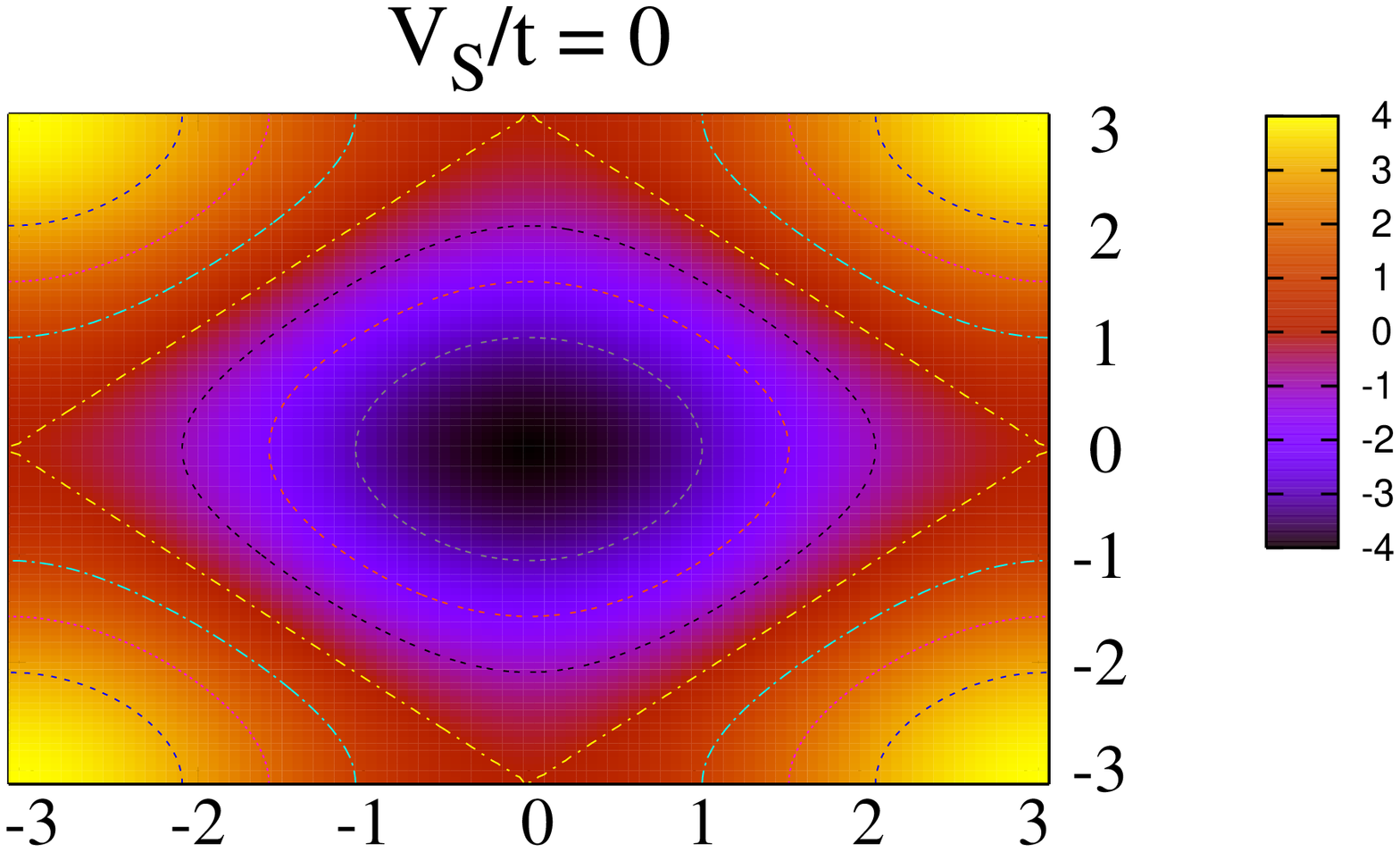}
\includegraphics[height=2.1in,width=2.1in,angle=-90]{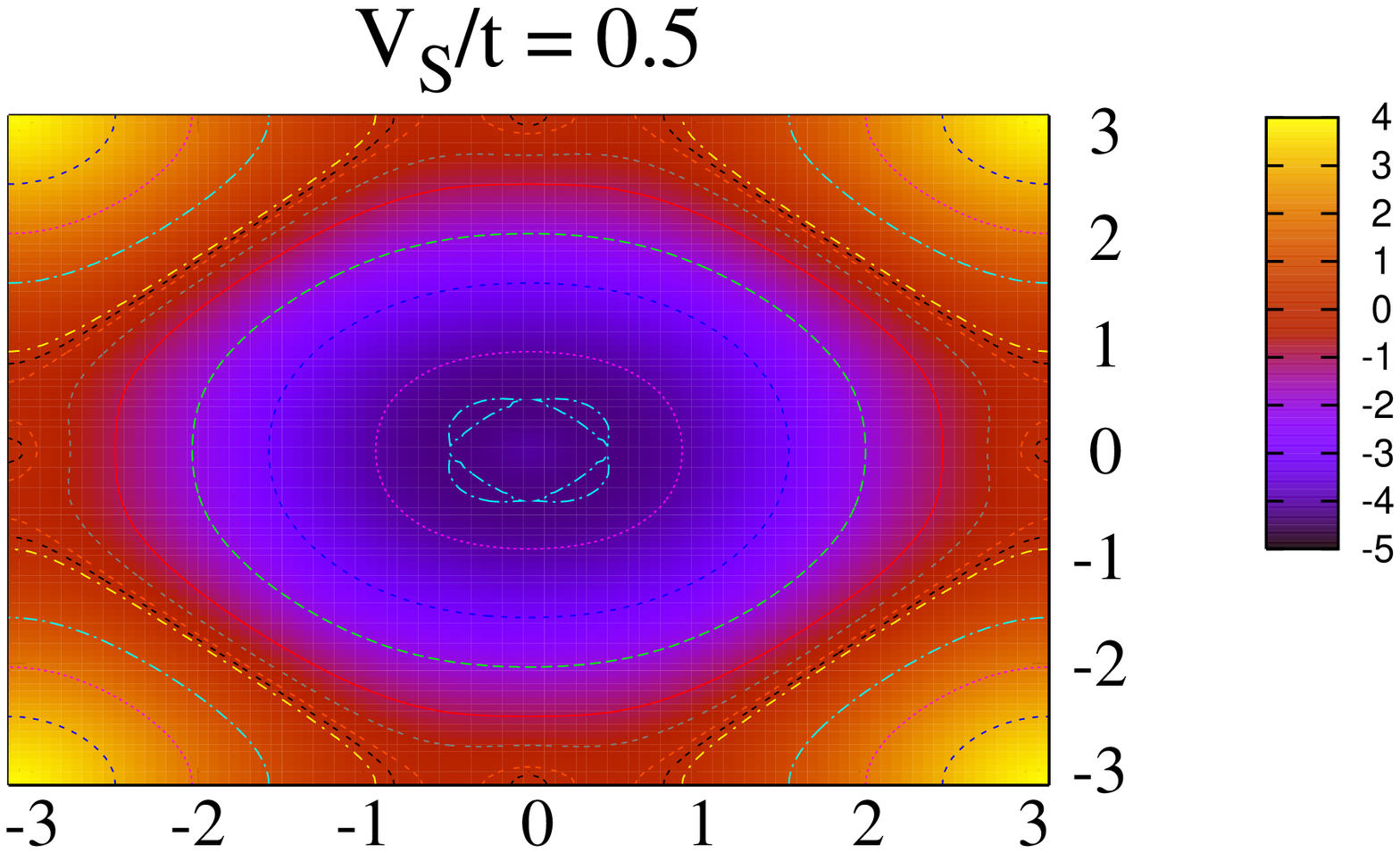}
\includegraphics[height=2.1in,width=2.1in,angle=-90]{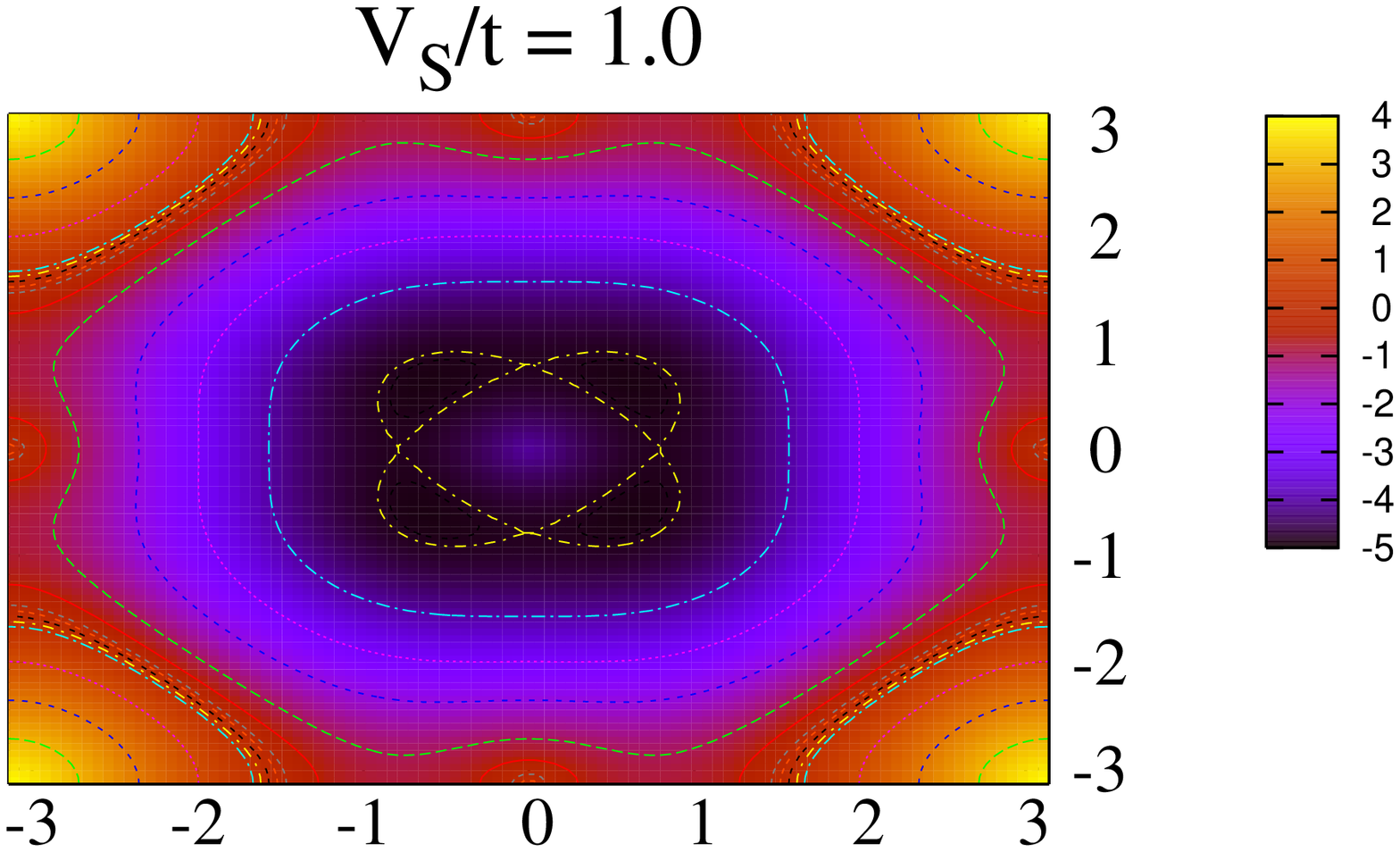}
\includegraphics[height=2.1in,width=2.1in,angle=-90]{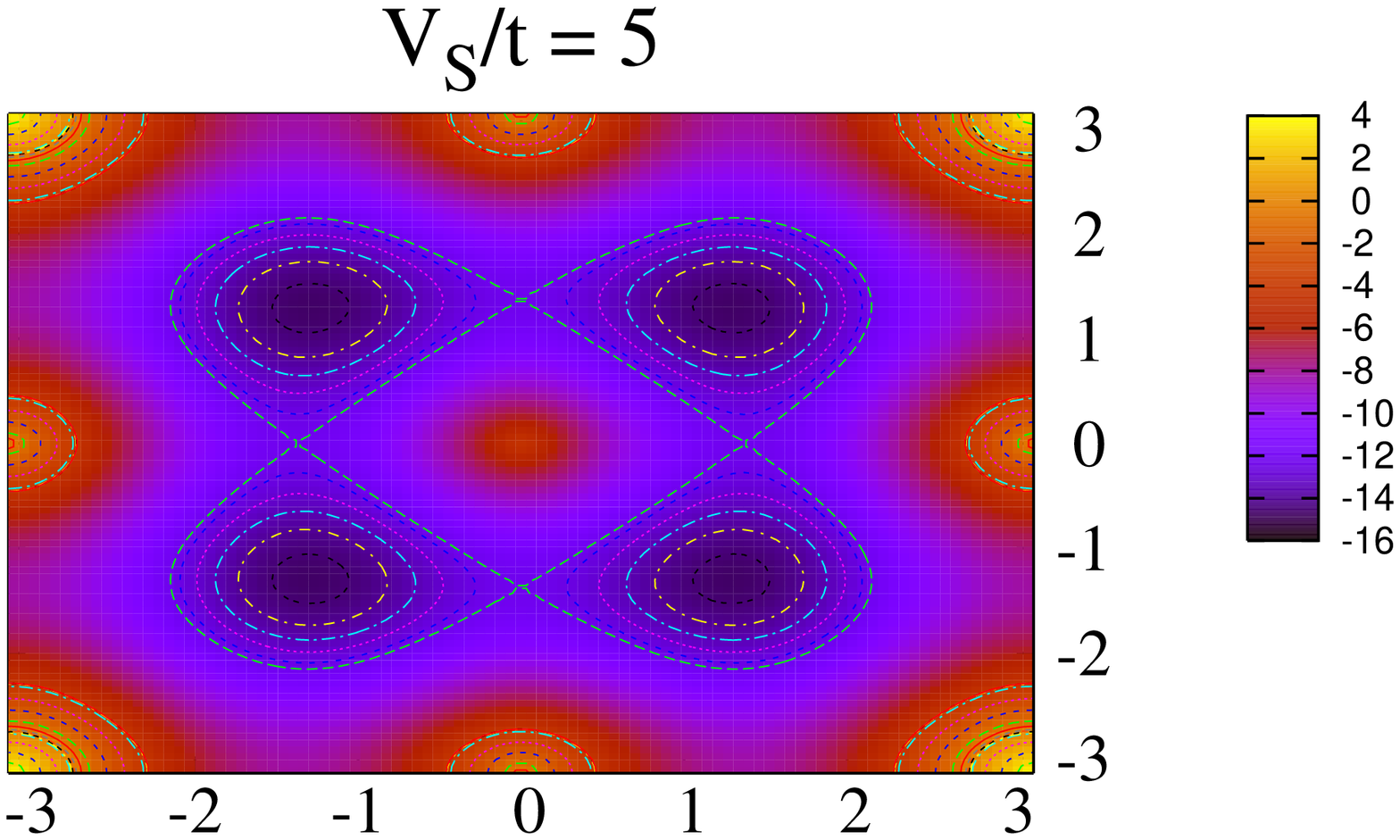}
\end{center}
\caption{ (color online) Contour plots for lower Rashba band with $%
V_{S}/t=0,0.5,1.0,5.0$. For $V_{S}=0$, there is only one energy minimum
point at $k_x=k_y=0$. For $V_{S}>0$, there are four energy minimum points
located at $k_{x}=k_{y}=\pm \arctan (\frac{V_{S}}{t\protect\sqrt{2}})$. For
nonzero $V_{S}$, there are also four saddle points near the energy minimum
points, which are located at $k_{x}=0,k_{y}=\pm \arctan (\frac{V_{S}}{t})$
and $k_{y}=0,k_{x}=\pm \arctan (\frac{V_{S}}{t})$. As $V_{S}$ increases, the
separation between minimum points and saddle points is increased (see
Fig.\protect\ref{fig2}(b)). }
\label{fig1}
\end{figure}

To examine this model in detail, we use a $2\times 2$ matrix to describe
the spin sector, and begin by excluding the phonon part of the Hamiltonian.
The remaining Hamiltonian is diagonalized through Bloch states in momentum space, written as
\begin{equation}
H_{0}=\sum_{\mathbf{k},\alpha }\epsilon _{\mathbf{k}}c_{\mathbf{k},\alpha
}^{\dag }c_{\mathbf{k},\alpha }+\sum_{\mathbf{k},\alpha ,\beta }\mathbf{%
\Omega }_{\mathbf{k}}\mathbf{\cdot \sigma }_{\alpha \beta }c_{\mathbf{k}%
,\alpha }^{\dag }c_{\mathbf{k},\beta }  \label{eqn1}
\end{equation}%
where $\epsilon _{\mathbf{k}}=-2t[\cos (k_{x})+\cos (k_{y})]$ and $\mathbf{%
\Omega }_{\mathbf{k}}\mathbf{\cdot \sigma }=$ $2V_{S}[\sin (k_{y})\sigma
_{x}-\sin (k_{x})\sigma _{y}]$ (we set the lattice spacing $a$ equal to unity).
Diagonalizing this $2\times 2$ matrix, we
get two bands, which we name the upper and lower Rashba bands. The
eigenvalues and eigenstates are given by
\begin{equation}
H_{0}\Psi _{\pm }=\varepsilon _{k,\pm }\Psi _{\pm },
\end{equation}%
with eigenvalues
\begin{equation}
\varepsilon _{k,\pm }=-2t[\cos (k_{x})+\cos (k_{y})]\pm 2V_{S}\sqrt{\sin
^{2}(k_{y})+\sin ^{2}(k_{x})},
\end{equation}%
and eigenvectors
\begin{equation}
\Psi _{\pm }=\frac{1}{\sqrt{2}}\left[ c_{k\uparrow }^{\dagger }\pm \frac{%
\sin (k_{y})-i\sin (k_{x})}{\sqrt{\sin ^{2}(k_{y})+\sin ^{2}(k_{x})}}%
c_{k\downarrow }^{\dagger }\right] |0\rangle .
\label{rashba_eigenstates}
\end{equation}%
In contrast to the model with parabolic bands, this model has a four-fold
degenerate ground state located at $k_{x}=k_{y}=\pm \arctan (\frac{V_{S}}{%
\sqrt{2}t})$,\cite{covaci09} which can be seen clearly from a contour plot of the lower
Rashba band in Fig.\ref{fig1}. There are also four saddle points near the
energy minimum points, which are located at $k_{x}=0,k_{y}=\pm \arctan (%
\frac{V_{S}}{t})$ and $k_{y}=0,k_{x}=\pm \arctan (\frac{V_{S}}{t})$. As $%
V_{S}$ increases, the separation between minimum points and saddle points is
enhanced (see below, in Fig.\ref{fig2}(b)).
The ground state energy for $H_{0}$ is given by $E_{0}=-4t\sqrt{%
1+V_{S}^{2}/(2t^{2})}$. Similarly, the effective mass along the diagonal is
\begin{equation}
\frac{m_{\mathrm{SO}}}{m_{0}}=\frac{1}{\sqrt{1+V_{S}^{2}/(2t^{2})}},
\label{eff_mass_so}
\end{equation}%
where $m_{0}\equiv 1/(4t)$ is the bare mass in the absence of spin-orbit
interaction, and $m_{\mathrm{SO}}$ is the effective mass due solely to
spin-orbit interaction. Note that the effective mass decreases due to
spin-orbit interaction. Below we will turn on the electron phonon
interaction, and the ground state energy (effective mass) will be further
lowered (raised) due to polaronic processes.

\subsection{non-interacting model: electron density of states}

\begin{figure}[tp]
\begin{center}
\includegraphics[height=2.5in,width=2.5in,angle=0]{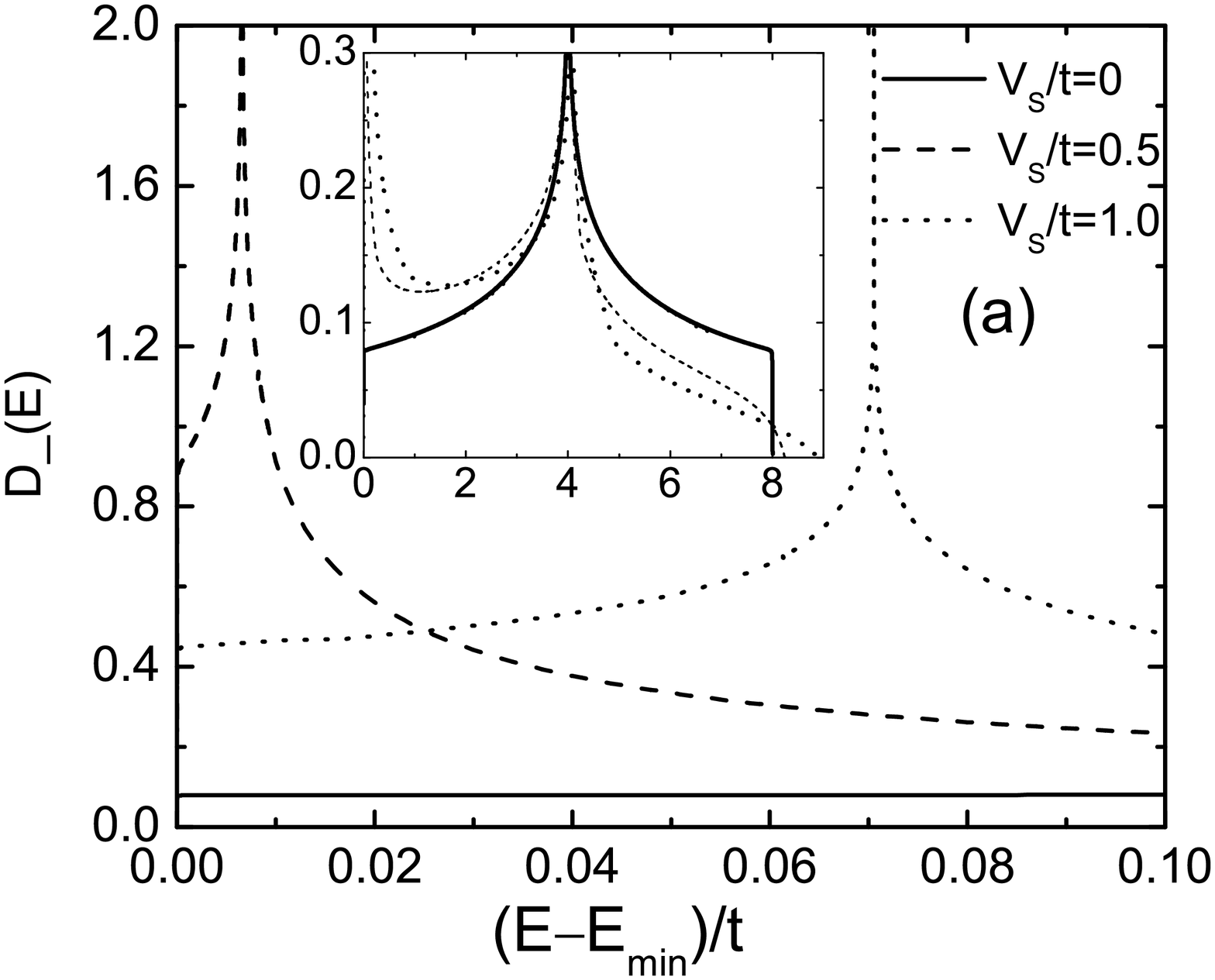} 
\includegraphics[height=2.5in,width=2.5in,angle=0]{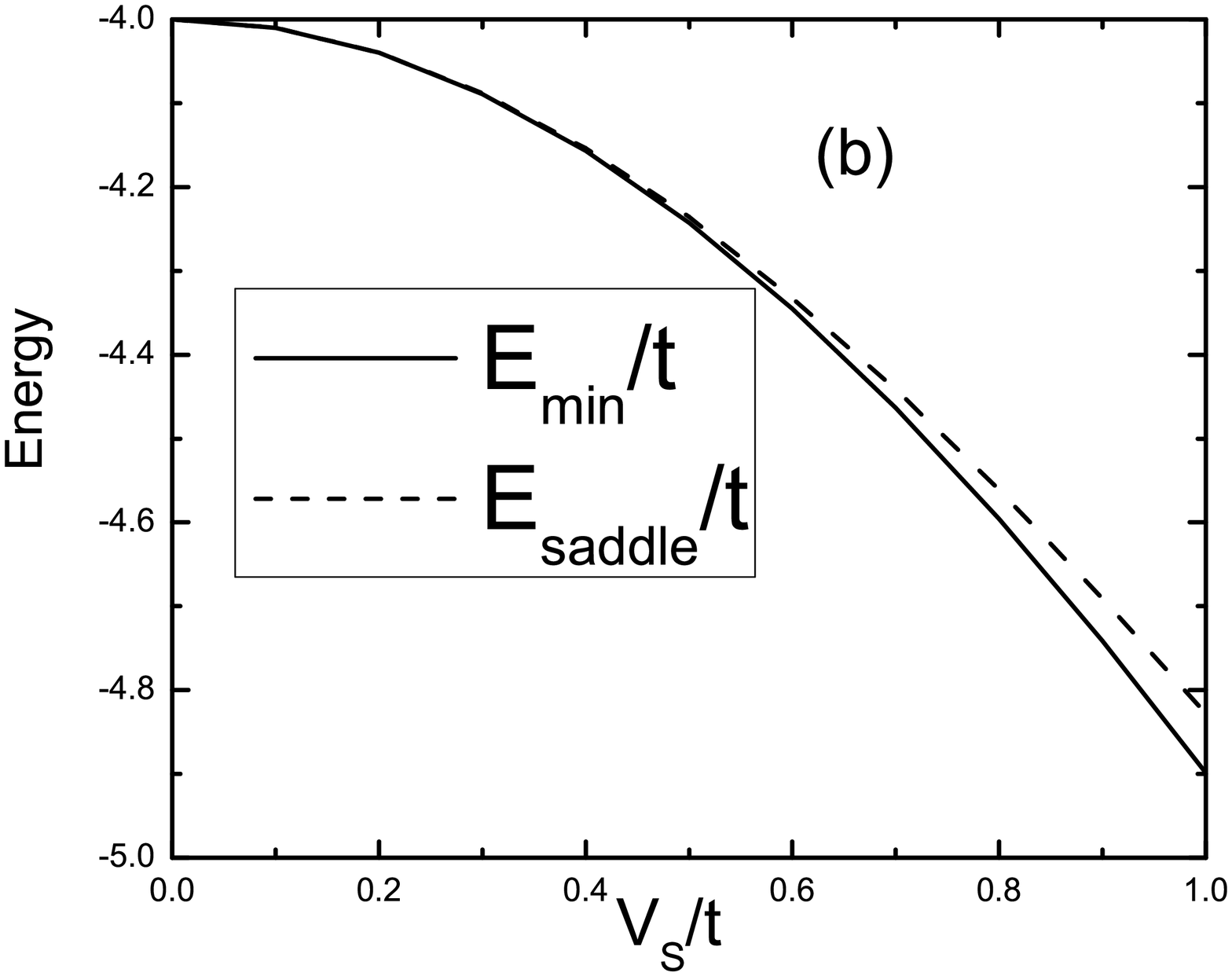}
\end{center}
\caption{(a) Non-interacting density of states $D_{-}(E)$ near the bottom of the
band for $V_{S}/t=0,0.5,1.0$. In the inset the density of states in the
whole band is shown for the same parameters. Note that the divergence at the
bottom of the band has been shifted to higher value.\protect\cite{covaci09}
(b) The separation between energy minimum points and saddle points as a
function of spin orbit interaction $V_{S}/t$. }
\label{fig2}
\end{figure}
The non interacting electron density of states (DOS) is defined for each band, as
\begin{equation}
D_s(E) = \sum_k \delta(E-\epsilon_{ks})
\label{dos}
\end{equation}
with $s =\pm 1$.

In the main frame of Fig.\ref{fig2}(a) we show the low energy DOS for various
values of the spin orbit interaction $V_S$; note that this involves
only $D_{-}(E)$ as the upper Rashba band exists only at higher energies. Furthermore,
information concerning the upper Rashba band can always be obtained through
the symmetry
\begin{equation}
D_+(E) = D_-(-E).
\label{sym}
\end{equation}
Fig.\ref{fig2}(a) shows that a divergence introduced by the spin orbit interaction
exists at higher energy,\cite{covaci09} and not at the bottom of the band, as occurs
for a parabolic dispersion.\cite{cappelluti07} This shift is due to the
separation of the energy minima from the saddle points in k-space, as shown in
Fig.\ref{fig2}(b). The saddle point energy is given by $E_{\rm sad} = -2t(1+\sqrt{1 + ({V_S/t})^2})$,
which is very close to the minimum energy $E_0$ even for sizeable $V_S/t$, as is evident from the
figure. This proximity of the divergence serves to elevate the value of the DOS at the bottom of the band.
With no spin orbit coupling this value is $D_{\pm}(E = E_0 = -4t) = 1/(4\pi t)$ ($V_S = 0$). With spin orbit coupling,
however, an expansion around the minimum energy $E_0 = -4t\sqrt{1 + V_S^2/2t^2}$ yields a DOS value
\begin{equation}
D_-(E=E_0) = {\sqrt{2} \over \pi}{1 \over V_S} \phantom{aaaaa}  V_S \ne 0.
\label{dos_at_bottom}
\end{equation}
Thus a discontinuity occurs as the spin orbit coupling is changed from zero --- the DOS immediately
has a divergence at the bottom of the band which, for any non-zero value of $V_S$, shifts to slightly
higher energy. The inset shows $D_-(E)$ over a wider energy range. Further details are provided in Appendix A.

\section{Ground state energy and effective mass}

When the electron phonon interaction is turned on, the ground state
energy (effective mass) will be lowered (increased) due to polaron effects. To
study the polaron problem numerically, we adopt a variational method
outlined by Trugman and coworkers,\cite{trugman90,bonca99} which could
determine polaron properties in the thermodynamic limit accurately. This
method was recently developed by Alvermann et al \cite{alvermann10} and Li
et al\cite{li10} to study the polaron problem near the adiabatic limit. In
this paper we will adopt the numerical techniques described in Ref.[\onlinecite{li10}].

\subsection{Strong coupling theory}

To investigate the strong coupling regime of the Rashba-Holstein model for a single polaron, we
use the Lang-Firsov\cite{lang63,marsiglio95} unitary transformation
$\overline{H}=e^{S}He^{-S}$ , where $S=g\sum_{i,\sigma }n_{i,\sigma
}(a_{i}-a_{i}^{\dagger })$, and obtain
\begin{equation}
\overline{H}=\overline{H}_{0}+\overline{T}
\end{equation}%
with
\begin{equation}
\overline{H}_{0} =\omega _{E}\sum_{i}a_{i}^{\dagger }a_{i}-g^{2}\omega
_{E}\sum_{i,\sigma }c_{i,\sigma }^{\dagger }c_{i,\sigma }
\label{ham0}
\end{equation}
and
\begin{eqnarray}
\overline{T}=&&-t\sum_{i,\sigma }(c_{i,\sigma }^{\dagger }c_{i+\hat{x},\sigma
}X_{i}^{\dagger }X_{i+\hat{x}}+c_{i,\sigma }^{\dagger }c_{i+\hat{y},\sigma
}X_{i}^{\dagger }X_{i+\hat{y}}+h.c.)  \notag \\
&&+ i V_{S}\sum_{i}(
c_{i,\alpha }^{\dagger }\sigma _{x}^{\alpha \beta }c_{i+\hat{y},\beta
}X_{i}^{\dagger }X_{i+\hat{y}} \nonumber \\
&&-c_{i,\alpha }^{\dagger }\sigma _{y}^{\alpha \beta }c_{i+\hat{x},\beta
}X_{i}^{\dagger }X_{i+\hat{x}} - h.c. ),
\end{eqnarray}%
where $X_{i}^{\dagger }=\exp \{g(a_{i}-a_{i}^{\dagger })\}$.\ Using
$e^{A+B}=e^{A}e^{B}e^{-1/2[A,B]}$, the hopping part of the Hamiltonian becomes
\begin{eqnarray}
\overline{T}=&&-te^{-g^{2}}\sum_{i,\sigma, \delta }\bigl[ c_{i,\sigma }^{\dagger }c_{i+\delta,\sigma}
(P_i^-)^\dagger (P_{i+\delta}^+)^\dagger P_{i}^+ P_{i+\delta}^- + h.c.\bigr] \notag \\
&&+iV_{S}e^{-g^{2}}\sum_{i}\bigl[
c_{i,\alpha }^{\dagger }\sigma _{x}^{\alpha \beta }c_{i+\hat{y},\beta}
(P_i^{-})^\dagger (P_{i+\hat{y}}^+)^\dagger P_i^+ P_{i+\hat{y}}^- \nonumber \\
&&-c_{i,\alpha }^{\dagger }\sigma _{y}^{\alpha \beta }c_{i+\hat{x},\beta}
(P_i^{-})^\dagger (P_{i+\hat{x}}^+)^\dagger P_i^+ P_{i+\hat{x}}^-  - h.c.\bigr],
\end{eqnarray}
where $P_i^{\pm} \equiv \exp{(\pm ga_{i})}$.
The unperturbed bare Hamiltonian, $\overline{H}_{0}$ provides the zeroth order energy for the polaron, and is already
diagonal for the single electron sector. The
eigenvalues are given by $E_{n}=n\omega _{E}-g^{2}\omega _{E},$ where $n$ is
the total number of phonons. Clearly the ground state has $n=0$, but remains
2N-fold degenerate, since the electron can occupy any one of the N sites and
it can have either spin up or spin down. If we consider the hopping term $\overline{T}$ as a
perturbation and apply degenerate perturbation theory to the 2N-fold degenerate
ground state, we need to diagonalize a $2N\times 2N$ matrix. A simpler approach is
to recognize that the
momentum $k$ is a good quantum number, and if we transform the original problem into
k-space, we need only solve a $2\times 2$ matrix which mixes the spin sectors; this results in
essentially Eq. (\ref{eqn1}), but with an extra band narrowing factor
$e^{-g^{2}}$. Thus we obtain the first order perturbation correction to the
energy as
\begin{equation}
E_{k \pm}= e^{-g^{2}} \varepsilon _{k \pm } - g^{2}\omega _{E},
\end{equation}
and the result is the familiar band narrowing factor that occurs when $V_S = 0$.

The eigenstates from degenerate perturbation theory are now simply Bloch-like states,
$\Psi _{\pm }$, as found in the non-interacting theory, Eq. (\ref{rashba_eigenstates}).
Thus the degeneracy is broken, and a comparatively narrower band is formed with
a minimum at a non-zero wave vector in the lower Rashba band, as found in the non-interacting
case.
To find the second order
correction to the ground state energy, we proceed as in Ref. [\onlinecite{marsiglio95}], and find
\begin{eqnarray}
&&E_{k - }^{(2)}= \sum_{n_{TOT}\neq 0,n_1,n_2,...=0,1,...\infty} \ \sum_{\ell = 1\atop \sigma}^N \nonumber \\
&&\frac{\left\vert \langle n_{1},n_{2},...n_{N}|_{ph}\otimes \langle c_{\ell \sigma}
|_{el} \overline{T} |\Psi _{k,-}\rangle _{el}\otimes |0\rangle _{ph}\right\vert ^2}{-n_{TOT}\omega _{E}}
\label{2nd_order}
\end{eqnarray}
where $n_{TOT}$ is the total number of phonons and $\Psi_{k,-}$ is given in Eq. (\ref{rashba_eigenstates}). 
With details shown in the appendix, we obtain
\begin{equation}
E_{k-}^{(2)} = -4 e^{-2g^2}\frac{t^2 + V_S^2}{\omega_E}\bigl[ f(2g^2) - f(g^2)\bigr] -e^{-2g^2}f(g^2)
\frac{\epsilon_{k-}^2}{\omega_E},
\label{2nd_order_b}
\end{equation}
where $f(x) \equiv \sum\limits_{n=1}^\infty
\frac{1}{n}\frac{x^{n}}{n!} \approx e^x/x \bigl[1 + 1/x + 2/x^2 +
  ...\bigr]$ 
(see Appendix). In some of the ensuing discussion, we will use the constant $\lambda$, familiar as the effective mass enhancement from weak coupling perturbation theory for the interacting electron gas. Here we use the definition
\cite{li10} 
$\lambda \equiv 2g^{2}\omega _{E}\frac{1}{4\pi t}$, since
$1/(4\pi t)$ is the value of the non-interacting electron density of states
for $V_{S}=0$ at the bottom of the band. Note that our definition of
$\lambda $ differs from that in Ref.[\onlinecite{covaci09}] or Ref.[\onlinecite{marsiglio95}]; both use the more conventional average density of states, $1/(8t)$.
Thus the ground state energy, excluding exponentially suppressed corrections, is
\begin{equation}
E_{GS}=-2\pi t\lambda\bigl( 1 + 2 \frac{t^2 + V_S^2}{(2\pi t \lambda)^2} \bigr),
\end{equation}
and there is a correction of order $1/\lambda^2$ compared to the zeroth order result.
Corrections in the dispersion enter in strong coupling only with an exponential supression.

\subsection{Weak coupling theory}

In the weak electron-phonon coupling regime, does spin-orbit coupling suppress or
enhance the "polaron effect" due to the electron-ion coupling? Weak coupling calculations
with a parabolic electron dispersion\cite{cappelluti07} showed an {\em increase} in the effective mass,
for example, as the spin-orbit coupling was increased. Here we perform weak coupling
perturbation theory, as described in Ref.[\onlinecite{cappelluti07}], with the same definitions,
except that the tight binding dispersion is used to describe the non-interacting electrons, as outlined
in the previous section. A straightforward calculation yields the self energy to first order in $\lambda$ as
\begin{equation}
\Sigma_{\rm weak}(\omega +i\delta )=\pi \lambda t\omega _{E}\sum_{\mathbf{%
k,}s=\pm }\frac{1}{\omega +i\delta -\omega _{E}-\varepsilon _{k,s}}.
\label{weak_a}
\end{equation}%
The effective mass can be obtained by the derivative of the self energy
\begin{equation}
\frac{m^{\ast }_{\rm weak}}{m_{SO}}=1-\frac{\partial }{\partial \omega }%
\Sigma_{\rm weak}(\omega +i\delta )|_{\omega =E_{0}}.
\label{weak_b}
\end{equation}

Near the adiabatic limit ($\omega _{E}\rightarrow 0$), by expanding $%
\varepsilon _{k,-}$ around $E_{0}$, as shown in the appendix for the
calculation of the DOS, we obtain

\begin{equation}
\frac{m^{\ast }_{\rm weak}}{m_{SO}}=1+\frac{\sqrt{2} \lambda t}{V_S},
\label{weak_mass}
\end{equation}
which shows a diverging effective mass  as the spin-orbit coupling increases. In fact, there is
a discontinuity for $V_S=0$, as the result is simply $\frac{m^{\ast }_{\rm weak}}{m_{SO}}=1+ \lambda/2$,
and $m_{SO} \rightarrow m_0 = 1/2t$, as given by Eq. (\ref{eff_mass_so}). Eq. (\ref{weak_mass}) will have a
limited domain of validity, however, as we will see below.
\begin{figure}[tp]
\begin{center}
\includegraphics[height=2.5in,width=2.5in,angle=0]{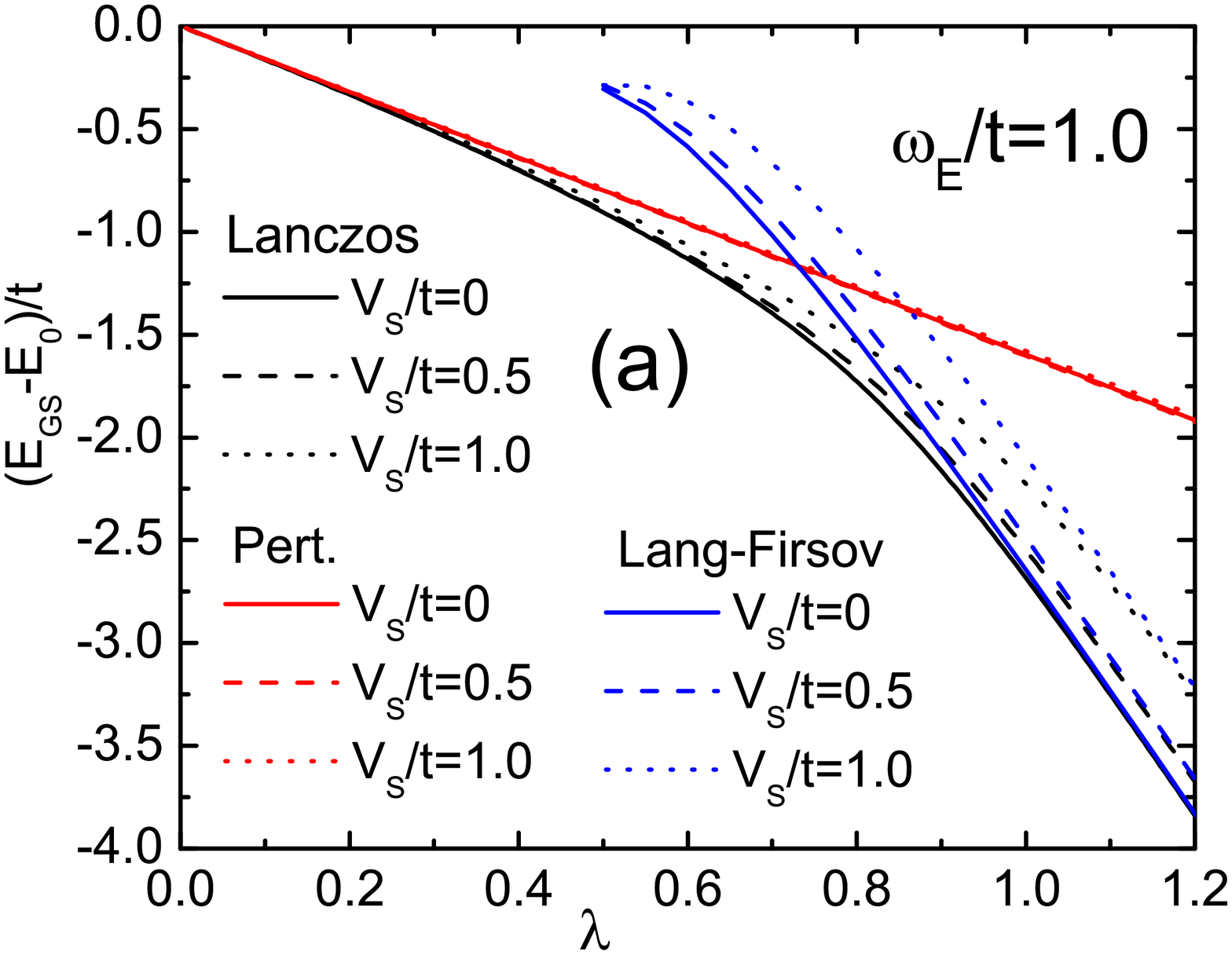}
\includegraphics[height=2.5in,width=2.5in,angle=0]{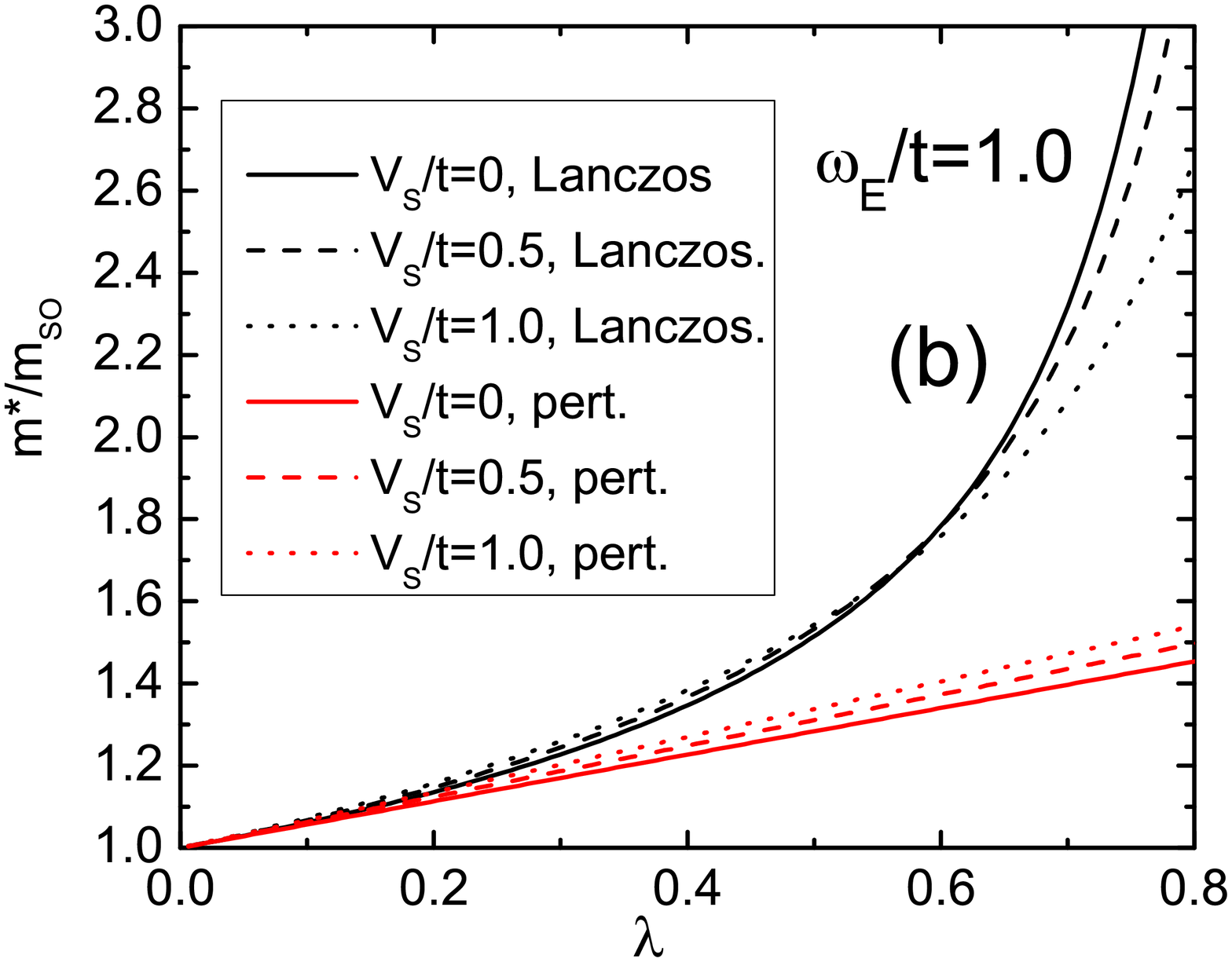}
\end{center}
\caption{(color online)(a) Ground state energy difference $E_{GS}-E_{0}$ vs. $\protect\lambda $
for $V_{S}/t=0,0.5,1.0$ and $\protect\omega _{E}/t=1.0$. Exact
numerical results are compared with those from weak coupling perturbation
theory (labeled "Pert." in the figure) and from Lang-Firsov strong coupling
theory. Agreement of both perturbative approaches with the exact numerical result is
excellent. The MA result (not shown) is also in excellent agreement with the numerical results.
(b)Effective mass $m^{\ast}/m_{SO}$ vs. $\protect\lambda $. Numerical
results are compared with that from weak coupling perturbation theory, and agreement is excellent for
low values of $\lambda$. Both exact and perturbative approaches show an enhanced effective mass
with increasing spin orbit coupling.}
\label{fig3}
\end{figure}

\subsection{Numerical Results}

In Fig. \ref{fig3}, we show the ground state energy and the effective mass
correction as a function of electron-phonon coupling $\lambda$, with non-zero values of the
spin orbit interaction, $V_{S}/t=0.5$ and $V_{S}/t=1.0$; these are
compared with the results from the Holstein model with $V_{S}/t=0.$ Here the
phonon frequency is set to be $\omega _{E}/t=1.0$, which is the typical
value used in Ref.[\onlinecite{covaci09}], and for each value of $V_S$,
the ground state energy is compared to the corresponding
result for $\lambda = 0$. The numerical results are
compared with results from weak coupling perturbation theory and  from Lang-Firsov strong
coupling theory.
\begin{figure}[tp]
\begin{center}
\includegraphics[height=2.5in,width=2.5in,angle=0]{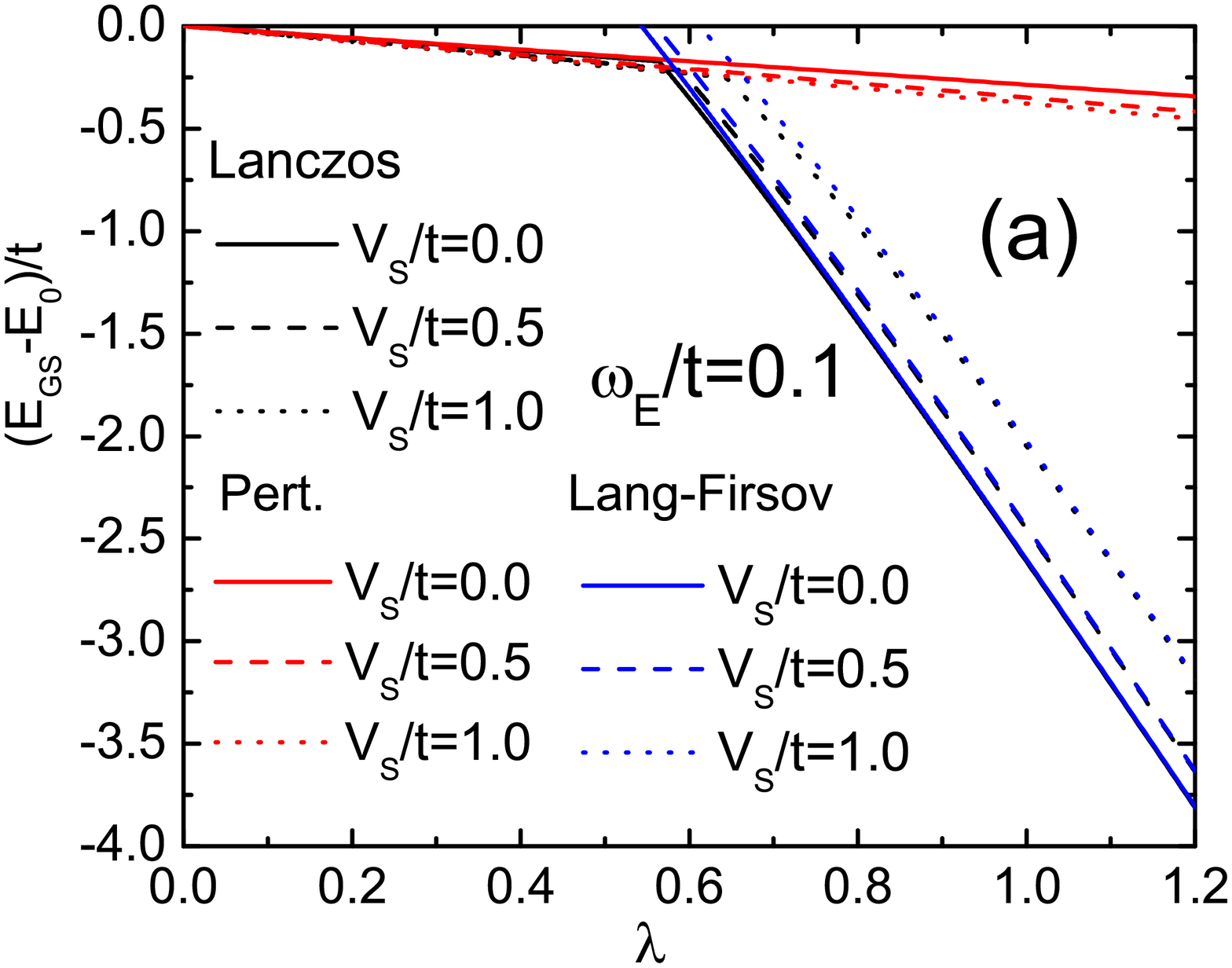} 
\includegraphics[height=2.5in,width=2.5in,angle=0]{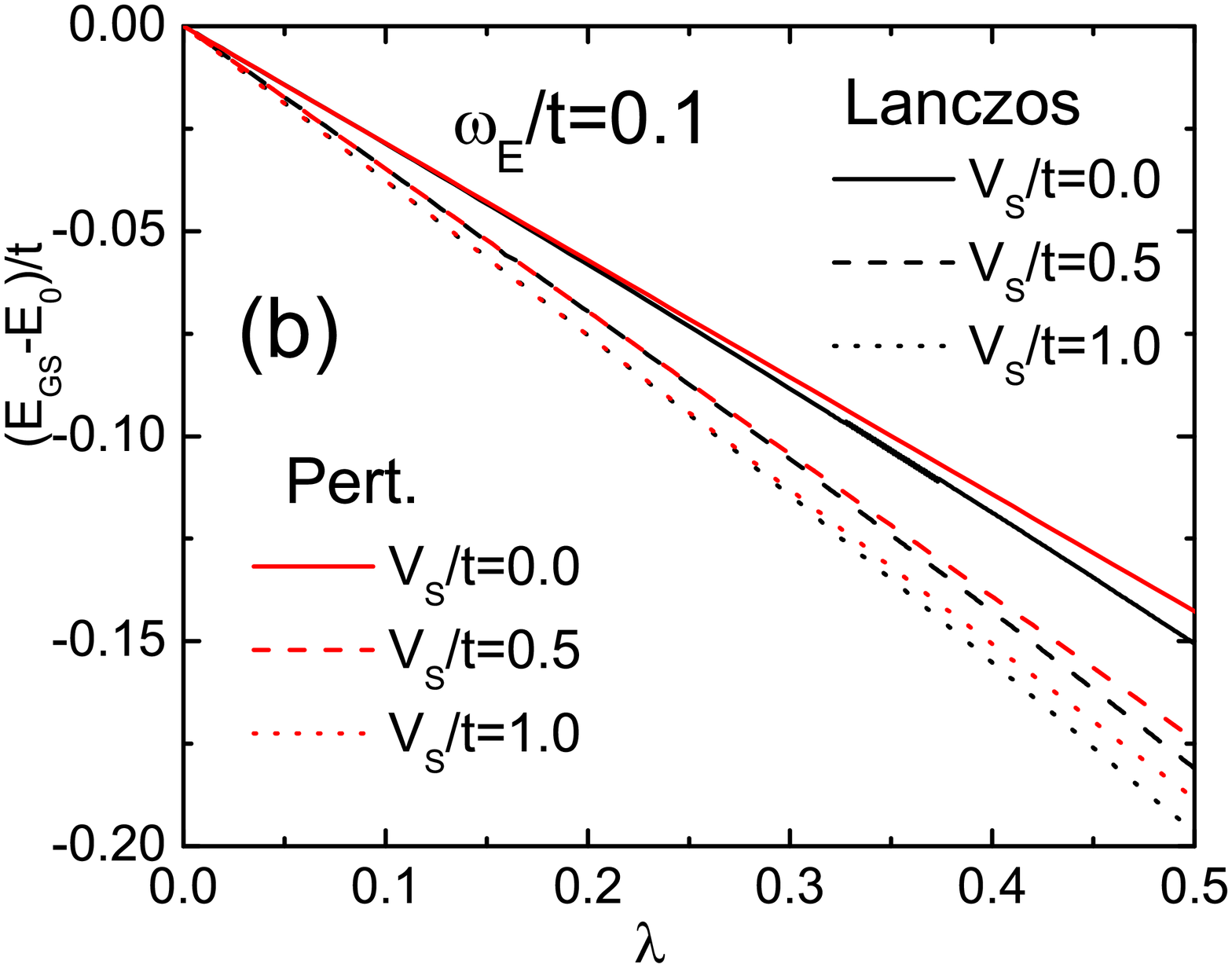} 
\includegraphics[height=2.5in,width=2.5in,angle=0]{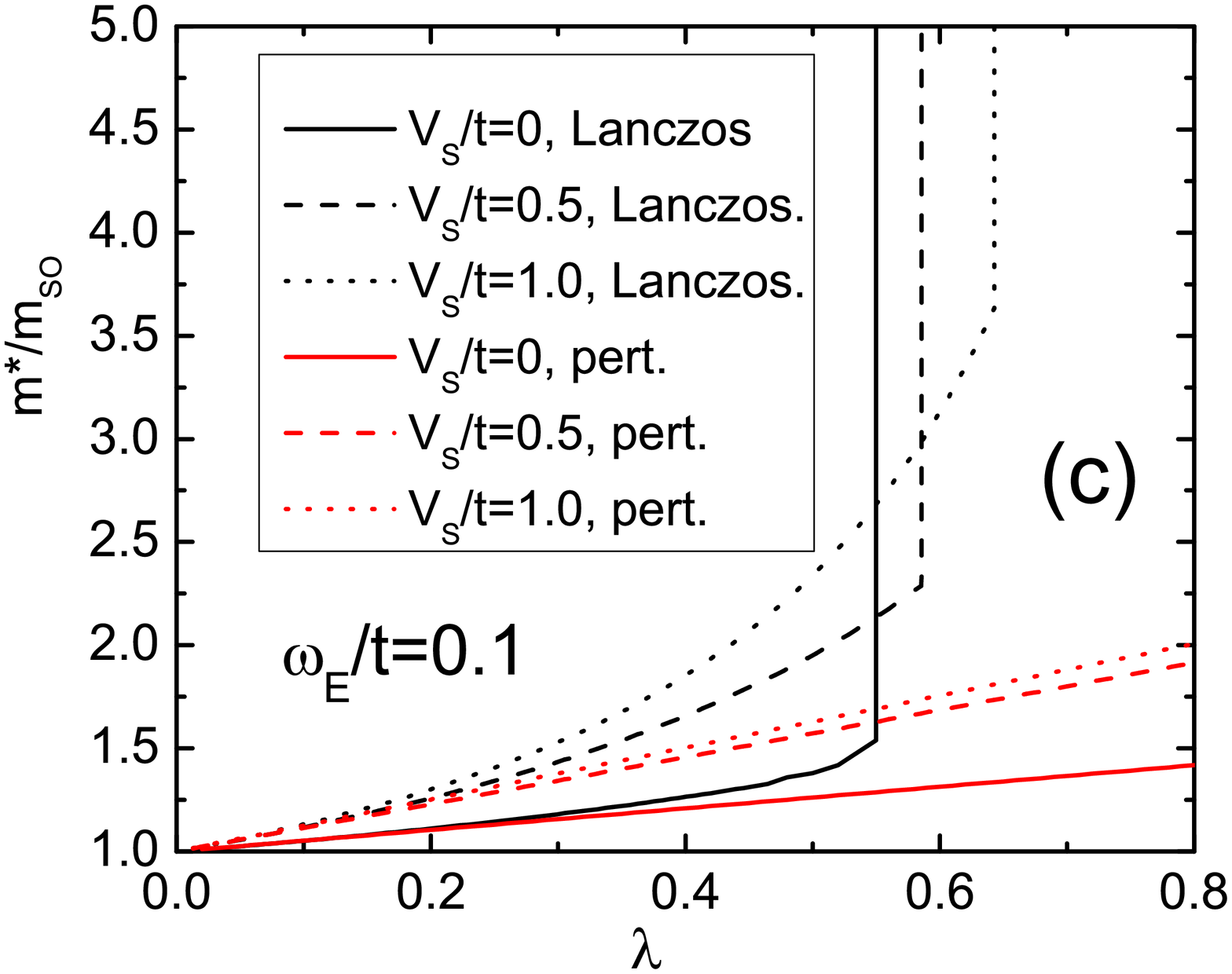}
\end{center}
\caption{(color online) (a) Ground state energy $E_{GS}-E_{0}$ vs. $\protect%
\lambda $ for $V_{S}/t=0,0.5,1.0$ and $\protect\omega _{E}/t=0.1$. Exact
numerical results are compared with those from weak coupling perturbation
theory (labeled Pert. in the fig) and Lang-Firsov strong coupling theory. (b)
Ground state energy $E_{GS}-E_{0}$ vs. $\protect\lambda $ in the weak and
intermediate coupling regime. (c) Effective mass $m^{\ast }/m_{SO}$ vs. $%
\protect\lambda $. Numerical results are compared with those from weak
coupling perturbation theory.}
\label{fig4}
\end{figure}
\begin{figure}[tp]
\begin{center}
\includegraphics[height=3.0in,width=3.0in,angle=0]{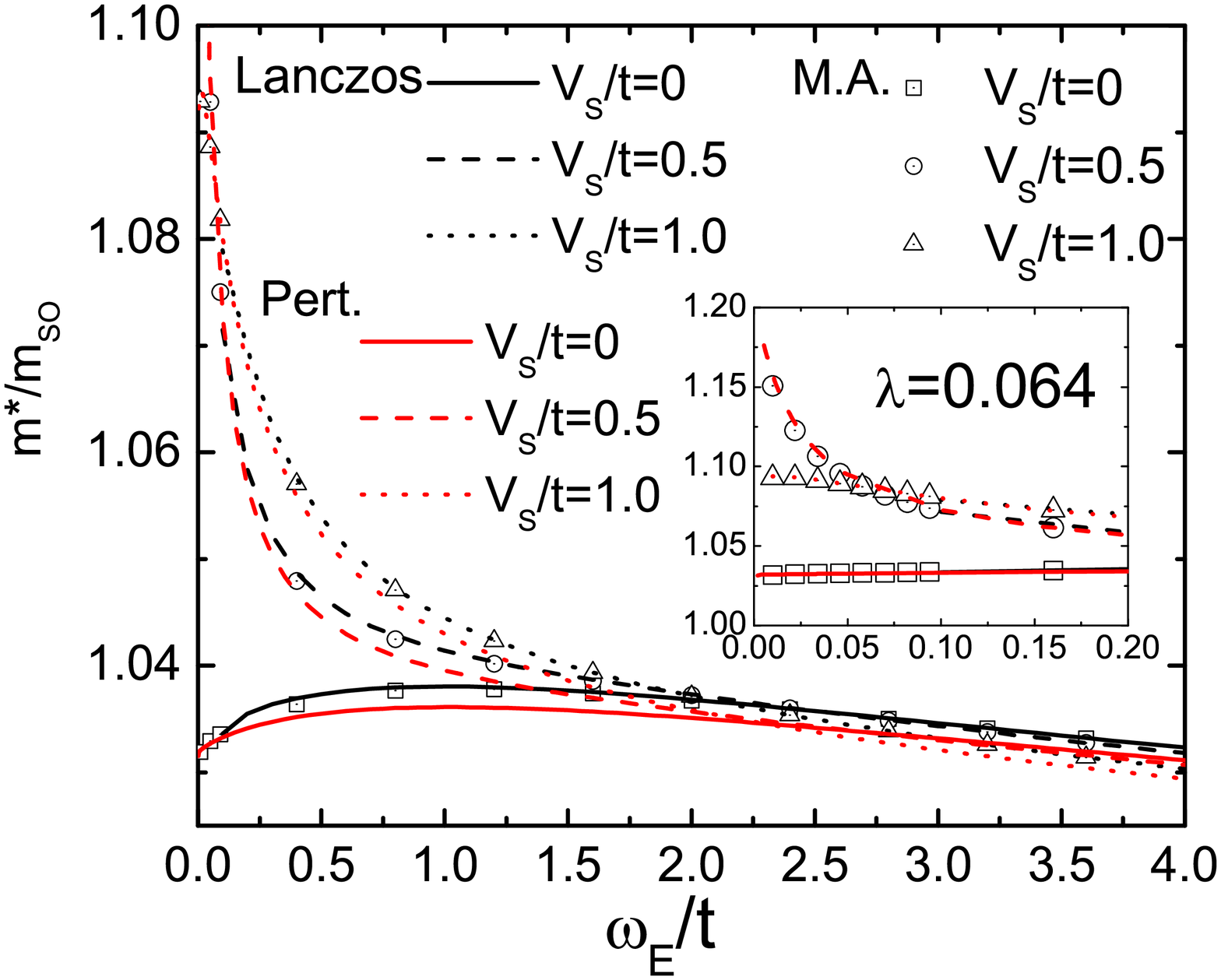}
\end{center}
\caption{(color online) Effective mass $m^{\ast }/m_{SO}$ vs. $\protect%
\omega _{E}/t$ for weak electron phonon coupling $\protect\lambda=0.064$. In
the inset the effective mass in the phonon frequency region near the
adiabatic limit is shown. It is clear that the effective mass is
enhanced as spin orbit interaction decreases near the adiabatic limit. This
is in agreement with the result inferred from the electron density of states shown in Fig.2(a). }
\label{fig5}
\end{figure}

In Fig.\ref{fig3}(a), the ground state energy crosses over
smoothly (at around $\lambda \approx 0.8$) from the delocalized electron regime to the small polaron regime.
Note that there is a slight dependence of the ground state energy on the spin orbit
interaction. If we define $\Delta E=E_{GS}-E_{0},$ then $\Delta
E(V_{S}/t=0.5)<\Delta E(V_{S}/t=0)<$ $\Delta E(V_{S}/t=1.0)\ $in the
delocalized electron regime, which is in agreement with the weak coupling
perturbation theory, though this is barely visible in the figure. In
the small polaron regime, the ground state energy is 
shifted up by the spin orbit interaction. This trend agrees with the results
from Lang-Firsov strong coupling theory. For $V_{S}/t=0,$ the Lang-Firsov theory agrees very
well with the numerical results, while as the spin orbit coupling $V_{S}$
increases, the Lang-Firsov theory becomes less accurate for the same electron
phonon coupling (e.g. if we look at $\lambda =1.0$, for $V_{S}/t=1.0,$ the
difference between Lang-Firsov theory and exact numerical results is larger
than that for $V_{S}/t=0$). This is due to the fact that the bandwidth is
increased by spin orbit interaction,\cite{covaci09} so the effective
electron phonon coupling is decreased by spin orbit interaction. Better
agreement with Lang-Firsov theory is achieved for larger values of $\lambda$.
In Fig.\ref{fig3}(b), the effective mass is enhanced by the spin orbit
interaction in the delocalized electron regime, which is in agreement with
the prediction from weak coupling perturbation theory. Here we have only shown results in
the region $V_{S}/t=0\sim 1.0$; for larger values of $V_{S}/t$ the effective mass will
be decreased by the spin orbit interaction in the delocalized regime.\cite{covaci09}
In the small polaron regime, the effective mass will always be
decreased by the spin orbit interaction.

In Fig.\ref{fig4}, we show the same results as Fig. \ref{fig3} for a
much smaller phonon frequency $\omega _{E}/t=0.1$, which is closer to the
adiabatic limit. In Fig.\ref{fig4}(a), the ground state
energy crosses over sharply (but still smoothly) from the delocalized electron regime to the small
polaron regime. If we use $\lambda_{c}$ to describe the critical value for
this sharp crossover, $\lambda _{c}$ will be enhanced significantly by the
spin orbit interaction. For $V_{S}/t=0.0,\lambda _{c}\simeq 0.55,$while for
$V_{S}/t=5.0,$ $\lambda _{c}\simeq 1.55$ from our numerical results. In Fig.
\ref{fig4}(b), in the delocalized electron regime the
ground state energy is decreased by the spin orbit interaction $\Delta
E(V_{S}/t=1.0)<\Delta E(V_{S}/t=0.5)<$ $\Delta E(V_{S}/t=0.0)$, which
is also in agreement with the weak coupling perturbation theory. For larger
$V_{S}/t$ the ground state energy will be increased\cite{covaci09}
in the delocalized electron regime. In the small polaron regime,
the ground state energy will be increased by the spin orbit interaction, in
agreement with the Lang-Firsov theory. In Fig.\ref{fig4}(c), the
effective mass enhancement for different spin orbit interaction $V_{S}/t$
is shown vs. electron phonon coupling strength, $\lambda$.
For $V_{S}/t=0$ there is a rather sharp crossover from the
delocalized electron regime to the small polaron regime.\cite{li10}

Near the crossover point, the effective mass enhancement for the delocalized
electron is around 1.4. For nonzero $V_{S}/t<1,$ near the crossover point,
the effective mass enhancement is higher, but still within the same
order of magnitute as $V_{S}/t=0$. This does not agree with the exact
adiabatic limit $\omega _{E}/t \equiv 0$ of the Rashba-Holstein model which has
been studied recently by Grimaldi,\cite{grimaldi10} based on a semiclassical method.
He found that for nonzero spin orbit interaction $V_{S}$, the ground
state will experience two phase transitions as the electron phonon coupling
$\lambda $ is increased. The first transition is from a delocalized electron to a
large polaron, while the second one is from a large polaron to a small polaron. In
Fig.\ref{fig4}(c), we observe only one sharp crossover from a
delocalized electron to a small polaron. Our results did not exclude the
possibilities that a large polaron regime will be found for $\omega
_{E}/t<0.1$, although we find this possibility unlikely. A similar circumstance holds in the absence of
a spin orbit coupling, where the adiabatic approximation gives rise to a single transition, while the quantum
calculations results only in a crossover. Smaller values of $\omega_E$ can be explored, but
quantum fluctuations become stronger for $\omega_{E}/t<0.1$ and the problem is numerically expensive for intermediate electron phonon coupling.

\begin{figure}[tp]
\begin{center}
\includegraphics[height=0.48\textwidth,angle=-90]{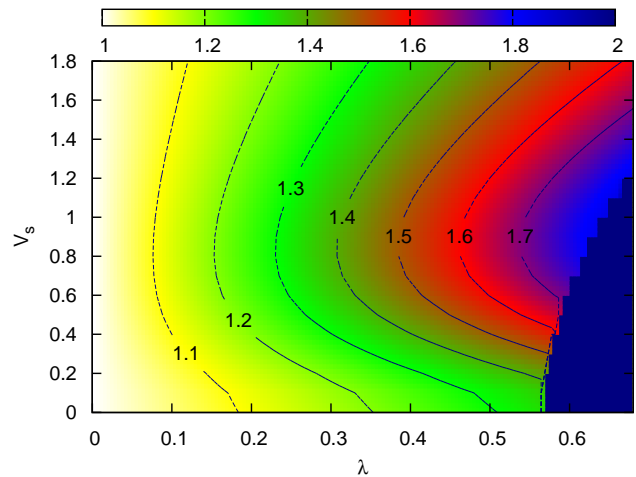}
\end{center}
\caption{(color online) Effective mass $m^{\ast }/m_{SO}$ map as a
function of spin orbit interaction $V_{S}/t$ and coupling constant
$\lambda$ for $\omega_E/t=0.1$ obtained with the momentum average
approximation.} 
\label{fig6}
\end{figure}

\begin{figure}[tp]
\begin{center}
\includegraphics[height=3.0in,width=3.0in,angle=0]{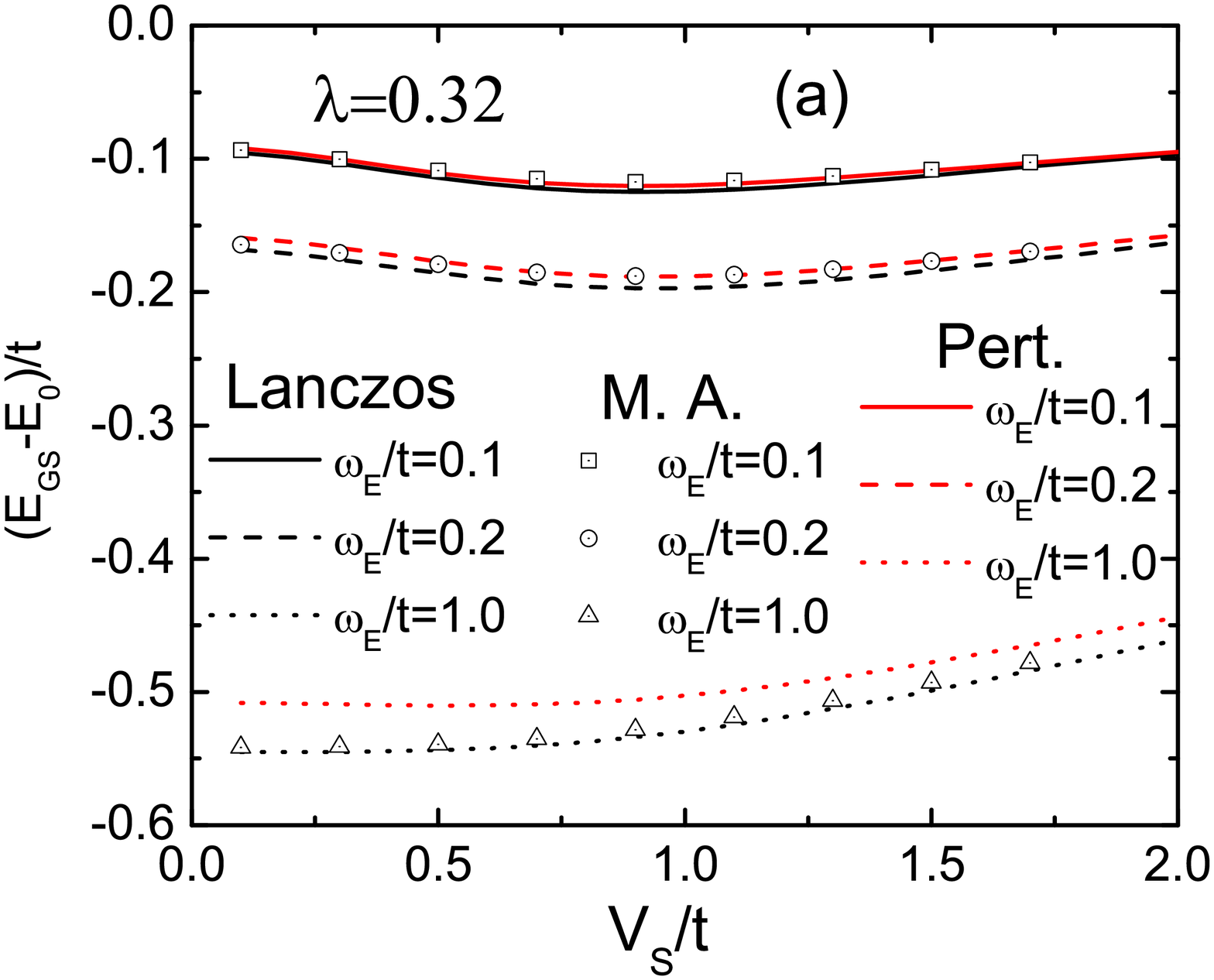} 
\includegraphics[height=3.0in,width=3.0in,angle=0]{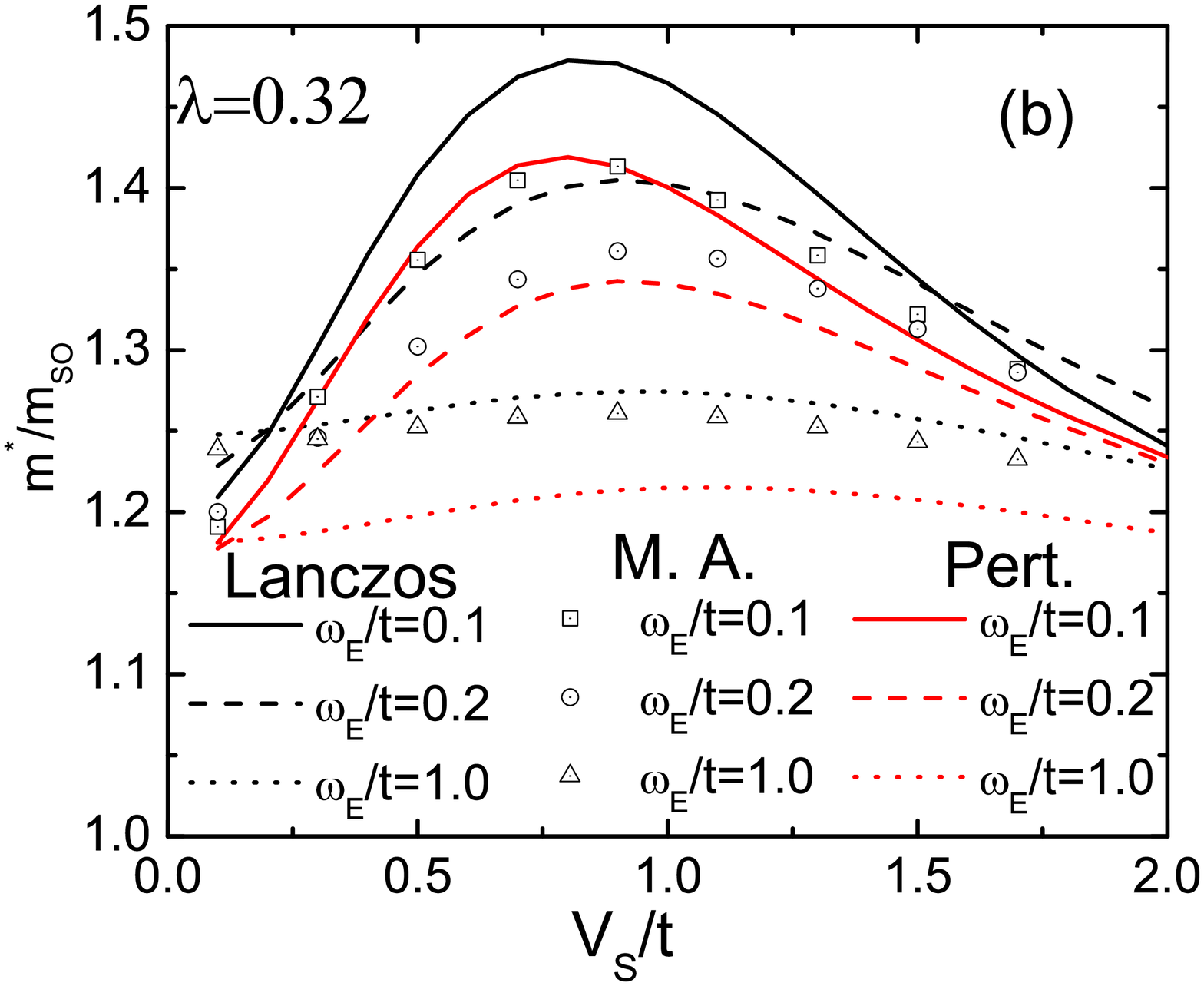}
\end{center}
\caption{(color online)(a) Ground state energy $E_{GS}-E_{0}$ as a function
of spin orbit interaction $V_{S}/t$ for $\protect\omega _{E}/t=0.1,0.2,1.0$
at $\protect\lambda =0.32$. (b) Effective mass $m^{\ast }/m_{SO}$ as a
function of spin orbit interaction $V_{S}/t$ for the same parameters. Exact
numerical results are compared with those from momentum averaging methods and
weak coupling perturbation theory. The Momentum Average approximation does not do as well
for low phonon frequencies.}
\label{fig7}
\end{figure}

To obtain some insight for the polaron effective mass near the adiabatic limit, we
resort to weak coupling perturbation theory. In Fig.\ref{fig5} we
observe an anomalous increase of the effective mass for small $\omega _{E}$
for nonzero $V_{S}$. However, the effective mass stops increasing as it
reaches some finite number (around $1.2$ and $1.1$ for $V_{S}/t=0.5$ and $1.0$, respectively), so
this does not indicate a breakdown of the perturbation theory. This result
is confirmed by the MA results, as illustrated. This is also in agreement with
results from the adiabatic limit. As shown in Fig.~2 of Ref.[\onlinecite{grimaldi10}],
 for $V_{S}/t=0.5$ and $1.0$ (his $\gamma /t=1.0$ and $2.0$),
the electron is definitely in the delocalized electron regime for $\lambda
=0.064$ ($\lambda =0.1$ in Ref.[\onlinecite{grimaldi10}]).
Actually this anomalous increase of effective mass is caused by an increase in the value of
the electron DOS at the bottom of the band, as shown in Fig.~2 and Eq. (\ref{dos_at_bottom}).
Thus, for even smaller values of $V_{S}/t,$ the anomalous mass enhancement will
increase further and perturbation theory will eventually break down. This is in agreement with the adiabatic
limit results --- as Fig.~2 of Ref.[\onlinecite{grimaldi10}] shows, for $V_{S}/t\simeq 0,$ the electron enters
the large polaron regime for small $\lambda $. As mentioned earlier, our results are consistent with
crossovers rather than transitions. This can be also seen in Fig.~\ref{fig6} where we plot for completeness a map of the effective mass as a function
of $V_s/t$ and $\lambda$ obtained by using the MA approximation for $\omega_E/t=0.1$. The exact results, while different in the details, show the same qualitative trends.

In Fig.~\ref{fig7}, we compare exact numerical results with both the
momentum average method\cite{covaci09} and with weak coupling perturbation theory,
for different values of $\omega _{E}.$ In Fig.~\ref{fig7}(a), the ground
state energy is shown as a function of $V_{S}/t$, while in Fig.~\ref{fig7}(b), the
effective mass is shown as a function of $V_{S}/t$. The MA
method agrees well with the exact numerical results for $\omega _{E}/t=1.0$.
For smaller values of $\omega _{E}$ ($\omega _{E}/t=0.1$ and $0.2$),
the Momentum Average approximation becomes less accurate and agrees more closely
with weak coupling perturbation theory. This is similar to what
happens for the Holstein model. Reasons for this quantitative
failure of MA in the adiabatic limit are explained in Ref. \onlinecite{berciu06}.

\section{Summary}

In this paper we have studied the problem of a single electron coupled to oscillating ions, in the presence
of a spin-orbit interaction. This problem has become relevant for a variety of spintronics applications. Many
previous treatments have addressed this problem with a finite density of electrons, and have therefore necessarily
required approximate theoretical methods for solution. The limit of
only one electron, previously solved with weak coupling perturbation
methods and with the momentum average approximation, is amenable to
exact solution as described here, and serves as a benchmark to which
other, approximate solutions must converge. Moreover, in many 
dilute semiconductor applications, the single electron result may be
the relevant regime required for understanding of the problem. 

The exact method of solution utilizes the Trugman method of
solution,\cite{bonca99} through Lanczos diagonalization. The procedure
for this is now well documented, and converges very quickly over a
very wide parameter regime. The momentum average
approximation\cite{covaci09} also works very well over the entire parameter
regime; there is a breakdown for very low phonon frequencies. In this
regime the adiabatic approximation\cite{grimaldi10} provides a good
qualitative picture. Weak coupling perturbation
theory\cite{cappelluti07} tends to be fairly accurate only for very
small coupling strengths. Finally, strong coupling perturbation
theory\cite{marsiglio95} is very accurate in the small polaron
regime. 

In weak coupling the presence of spin orbit coupling increases the effective mass of the electron coupled to
Einstein phonons.\cite{cappelluti07} The effective mass is small to
begin with, so in this regime the impact of spin orbit coupling is
fairly minor. As the electron phonon coupling increases, and one
enters the small polaron regime, the presence of spin orbit coupling
has the opposite effect, as first noted with the momentum average
approximation.\cite{covaci09} Since in this regime the effective
masses can be quite large, spin orbit coupling can have a profound
effect on the characteristics of the electron. 

\begin{acknowledgments}

This work was supported in part by the Natural Sciences and Engineering
Research Council of Canada (NSERC), by ICORE (Alberta), by Alberta Ingenuity, by the Flemish Science Foundation (FWO-Vl) and by the Canadian
Institute for Advanced Research (CIfAR).
\end{acknowledgments}

\appendix

\section{Density of states at the bottom of the band}

Expanding $\varepsilon _{k,-}$ around the minimum energy $E_{0}$, by
defining $k_{x}^{\prime }=k_{x}\pm \arctan (\frac{V_{S}}{\sqrt{2}t}%
),k_{y}^{\prime }=k_{y}\pm \arctan (\frac{V_{S}}{\sqrt{2}t}),$ we have
\begin{equation}
\varepsilon _{k,-}-E_{0}=\frac{0.5t}{\sqrt{1+V_{S}^{2}/(2t^{2})}}%
[(1+V_{S}^{2}/t^{2})(k_{x}^{\prime 2}+k_{y}^{\prime 2})\pm 2k_{x}^{\prime
}k_{y}^{\prime }]
\label{expand}
\end{equation}%
To calculate the density of states at the bottom of the band, from the
definition, we have
\begin{equation}
D_-(E_{0}+E_{1})=\frac{1}{4\pi ^{2}}\int_{-\pi }^{\pi
}dk_{x}\int_{-\pi }^{\pi }dk_{y}\delta (E_{0}+E_{1}-\varepsilon _{k,-}),
\end{equation}%
where $E_{1}$ is a small amount of energy above the bottom of the band, $E_{0}$. 
Around the four energy minimum points there
are four small regions which will contribute to this integral. We choose one
of them (and then times our results by a factor of 4) and use the
definitions of $k^{\prime }$ above instead of $k,$ introduce a small cutoff $%
k_{c}$ which is the radius of a small circle around $k_{\min },$ thus the
integral reads
\begin{eqnarray}
&&D_-(E_{0}+E_{1})=4\times \frac{1}{4\pi ^{2}}\int_{0}^{k_{c}}k^{%
\prime }dk^{\prime }\int_{-\pi }^{\pi }d\theta  \notag \\
&&\times \delta (E_{1}-\frac{0.5t}{\sqrt{1+V_{S}^{2}/(2t^{2})}}%
[(1+V_{S}^{2}/t^{2})+\sin 2\theta ]k^{\prime 2})  \notag \\
&=&\frac{\sqrt{1+V_{S}^{2}/(2t^{2})}}{\pi ^{2}t}\int_{-\pi }^{\pi }d\theta
\frac{1}{[(1+V_{S}^{2}/t^{2})+\sin 2\theta ]}  \notag \\
&=&{\sqrt{2} \over \pi}{1 \over V_S}.
\end{eqnarray}

The derivation of the effective mass in the weak coupling approximation (Eq. (\ref{weak_mass})) proceeds
similarly.
We begin with Eq. (\ref{weak_a}) in the text for the self energy. For very small phonon frequency we need only
focus on the lower Rashba band, $s=-1$. Furthermore, the non-interacting electron energy can be expanded
about a minimum, as in Eq. (\ref{expand}). Noting that there are four equal contributions coming from the four
degenerate minima, we obtain
\begin{eqnarray}
&&\Sigma_{\rm weak}(\omega + i\delta) = -4 \frac{\pi \lambda t \omega_E}{(2\pi)^2} \int dk_x^\prime \int dk_y^\prime \nonumber \\
&& \frac{1}{a^2 + \frac{t}{2\sqrt{1+{V_S^2 \over 2 t^2}}}\biggl[ (1+(V_S/t)^2)(k_x^{\prime 2} +k_y^{\prime 2}) + 2k_x^\prime k_y^\prime \biggr] }
\label{a4}
\end{eqnarray}
where $a^2 = E_0 + \omega_E - \omega$, and the integration is understood to be around a small disk located at one of the energy minima. Transforming to polar coordinates allows both the radial and angular integral to be done analytically; for the radial integral we keep only the dominant portion for small $\omega_E$, and, after differentiation, we readily obtain the result quoted in the text (Eq. (\ref{weak_mass})).

\section{Strong coupling limit}

To investigate the strong coupling limit using second order perturbation, we
need to evaluate Eq. (\ref{2nd_order}), repeated here for convenience:
\begin{eqnarray}
&&E_{k - }^{(2)}= \sum_{n_{TOT}\neq 0,n1,n2,...=0,1,...\infty} \ \sum_{\ell = 1\atop \sigma}^N \nonumber \\
&&\frac{\left\vert \langle n_{1},n_{2},...n_{N}|_{ph}\otimes \langle c_{\ell \sigma}
|_{el} \overline{T} |\Psi _{k,-}\rangle _{el}\otimes |0\rangle _{ph}\right\vert ^2}{-n_{TOT}\omega _{E}} \nonumber \\
&& = \frac{-t^2e^{-2g^2}}{\omega_E} \sum_{{n1,n2..=0}\atop {n_{\rm TOT}\neq 0}}^\infty \ \sum_{\ell = 1}^N
\frac{|A_{\uparrow}|^2 + |A_{\downarrow}|^2}{n_{\rm TOT}},
\label{a:2nd_order}
\end{eqnarray}
where $A_\sigma$ is given a series of matrix elements (distinct for $\sigma = \uparrow$ and $\downarrow$).
These turn out to give equal contributions, so we illustrate in some detail the result for $A_\uparrow$ only.
After some algebra, we obtain
\begin{equation}
|A_\uparrow|^2 = |u_\ell(-g)|^2 \ |\sum_{\delta=\pm x, \pm y}c_\delta u_{\ell+\delta}(g)|^2,
\label{b2}
\end{equation}
where
\begin{equation}
u_\ell(\pm g) \equiv \langle n_\ell | e^{\pm ga_\ell^\dagger}|0 \rangle = \frac{(\pm g)^{n_\ell}}{\sqrt{n_\ell !}}
\label{b3}
\end{equation}
and
\begin{eqnarray}
c_{+x} &=& e^{+ik_xa}\bigl(1 + \frac{V_S}{t}e^{i\phi_k} \bigr) \nonumber \\
c_{-x} &=& e^{-ik_xa}\bigl(1 - \frac{V_S}{t}e^{i\phi_k} \bigr) \nonumber \\
c_{+y} &=& e^{+ik_ya}\bigl(1 - i\frac{V_S}{t}e^{i\phi_k} \bigr) \nonumber \\
c_{-y} &=& e^{-ik_ya}\bigl(1 + i\frac{V_S}{t}e^{i\phi_k} \bigr),
\label{b4}
\end{eqnarray}
and
\begin{equation}
e^{i\phi_k} \equiv \frac{\sin{(k_y a)} - i\sin{(k_x a)}}{\sqrt{\sin^2{(k_x a)} + \sin^2{(k_y a)}}}.
\label{phi}
\end{equation}
For each of the $u_\ell(\pm g)$ in Eq. (\ref{b2}) it is to be understood that $n_\ell \ne 0$, but all other
$n_{\ell{^\prime}} = 0$ for $\ell^\prime \ne \ell$. Hence, in the 16 terms in Eq. (\ref{b2}), 12 will
have all phonon numbers equal to zero (other than $n_\ell$); the other 4 will have both $n_\ell$ and $n_{\ell + x}$
(or  $n_\ell$ and $n_{\ell - x}$, etc.)
not equal to zero in general. As already mentioned, the contribution from $|A_\downarrow|^2$ is identical to that from
$|A_\uparrow|^2$, so this merely gives us a factor of $2$ in Eq. (\ref{a:2nd_order}). Moreover, translational invariance makes the contribution from each site identical, so the sum over sites is trivially performed. This equation then becomes
\begin{equation}
E_{k-}^{(2)} = -\frac{4t^2 e^{-2g^2}}{\omega_E} \{f(g^2)\bigl(\frac{\epsilon_{k-}}{2t}\bigr)^2 + \bigl[f(2g^2) - f(g^2)\bigr] \bigl[1 + (\frac{V_S}{t})^2\bigr] \},
\label{b6}
\end{equation}
where
\begin{eqnarray}
f(x) &\equiv& \sum\limits_{n=1}^\infty \frac{1}{n}\frac{x^{n}}{n!}  = Ei(x) - \gamma - \ln{x} \nonumber \\
&\approx &e^x/x \bigl[1 + 1/x + 2/x^2 + ...\bigr],
\label{fx}
\end{eqnarray}
and $Ei(x)$ is the exponential integral and $\gamma \approx 0.5772$ is Euler's constant. Eq. (\ref{b6})
leads directly to Eq. (\ref{2nd_order_b}) in the text.

$^*$ present address: Dept. of Oncology, University of Alberta, Edmonton, AB, Canada T6G 1Z2 \\

\end{document}